\documentclass[11pt]{article}
\usepackage{vietnam,epsfig}
\usepackage{amsmath}
\usepackage{amscd}
\usepackage{amssymb}

\def\Journal#1#2#3#4{{#1} {\bf #2}, #3 (#4)}

\def\NPB{{\em Nucl. Phys.} B}
\def\PLB{{\em Phys. Lett.}  B}
\def\PRL{\em Phys. Rev. Lett.}
\def\PRD{{\em Phys. Rev.} D}

\def\EPJ{{\em Eur. Phys. J.} C}
\def\CPC{\em Comp. Phys. Comm.}

\def\be{\begin{equation}}
\def\ee{\end{equation}}
\def\bea{\begin{eqnarray}}
\def\eea{\end{eqnarray}}

\newcommand{\ke}{$\rm K^{0}_{e3}$}
   
\newcommand{\kel}{$\rm K_{L}\rightarrow~\pi^{\pm}~e^{\mp}~\nu_{e}$}
\newcommand{\keg}{$\rm K^{0}_{e3 \gamma}$}
\newcommand{\kegl}{$\rm K_{L}\rightarrow~\pi^{\pm}~e^{\mp}~\nu_{e} \gamma$}
\newcommand{\kl}{$\rm K_{L} $}
\newcommand{\vus}{{$\vert V_{{\it us}} \vert$}}
\newcommand{\bmath}{\begin{displaymath}}
\newcommand{\emath}{\end{displaymath}}

\begin{document}
\vspace*{4cm}
\title{SEMILEPTONIC $\rm K_{L}$ DECAYS AT NA48}

\author{ M. VELTRI }

\address{Istituto di Fisica, Universit\`a di Urbino and INFN Sezione 
Firenze, Via S. Chiara 27,\\
I--61029 Urbino, Italy}

\maketitle\abstracts{
\vskip -1 cm
Preliminary results on $K_{L}\rightarrow \pi e \nu (\gamma)$ decays
collected by the NA48 detector at the CERN SPS are reported.
Using a sample of $6.8\times10^{6}$ reconstructed events
$BR($ \ke $) = 0.4010\pm 0.0028 \pm0.0035$ was obtained. From the branching 
ratio the value of \vus$ = 0.2187 \pm 0.0028 $ was extracted.
The same data sample has provided also a high precision measurement
of the slope $\lambda_{+}$ of the form factor of the \ke~decay. 
Investigating the \keg~decay, from a sample of 18977 reconstructed events, 
$BR($ \keg $)/ BR($ \ke $) = (0.964\pm0.008^{+0.012}_{-0.011})\%$ 
was determined. }

\section{Introduction}\label{sec:intro}
The NA48 experiment at the CERN SPS has been exploited in a rich
program of kaon physics. The existence of direct CP violation
in the decay of two pions was demonstrated and many rare kaon decay 
searches as well as high precision measurements have been carried out.\\
Here will be reported three recent measurements of semileptonic \kl~decays,
namely the \ke~branching ratio together with a determination of the CKM 
matrix element \vus,  the \ke~form factors and the \keg~branching ratio.\\
The most relevant component of the NA48 detector~\cite{na48det}
are a high resolution liquid--krypton electromagnetic calorimeter
and a magnetic spectrometer consisting of 4 drift chambers and a dipole
magnet located inside a helium tank. Other components are a hodoscope
for precise track time determination, a hadronic calorimeter and a muon 
veto.\\
The data sample used for all the three analyses was taken in 1999 during a 
dedicated minumum bias run without $K_{S}$ beam; 
similar selection criteria were adopted throughout
in order to select the events.

\section{Physics Motivations for the Study of Kaon 
Semileptonic Decays}\label{sec:ke3phys}
The renewed interest in the study of the semiletponic \kl~decays
is related to the long standing problem about the unitarity of 
the CKM matrix. The unitarity of the CKM matrix requires for the 
first row that:\\
\be
   \vert V_{{\it ud}} \vert^2~+~\vert V_{{\it us}} \vert^2
~+~\vert V_{{\it ub}} \vert^2 ~=~ 1
\ee
a significant departure from unity in this relation would imply the 
breaking of unitarity and the existence of new physics.
PDG 2004~\cite{pdg2004} values indicates a $2.2~\sigma$ deviation from 
unity and \vus~is responsible for about 50\% of the error.
\vus~can be extracted from the kaon semileptonic decays
($K_{\ell 3}, \ell=e, \mu$) width via the relation:
\be
\Gamma_{K \ell 3} ~=~ \frac { G^{2}_{F} M^{5}_{K} } { 192 \pi^{3} }~
~S_{EW}(1+\delta^{\ell}_{K})~ C^2~ \vert  V_{{\it us}} \vert^2 ~
f^{2}_{+}(0) ~ I^{\ell}_{K}(\lambda_{+},\lambda_{0})
\ee

\noindent where $G_{F}$ is the Fermi coupling constant, $ M_{K}$
the kaon mass, $S_{EW}$ and $\delta^{\ell}_{K}$ the short and the 
long distance radiative corrections, $C^{2}$ is 1 
for $K^{0}_{L}$  and 1/2 for $K^{\pm}$ decays,
$f_{+}(0)$ is the calculated form factor at zero momentum transfer and 
finally $I^{\ell}_{K}(\lambda_{+},\lambda_{0})$ is the phase space 
integral which depends on the two form factors which describe the decay.
Being pure vector transitions the $K_{\ell 3}$ decays suffer only 
from second order corrections in the simmetry breaking and therefore
are the most accurate and theoretically cleanest way to extract \vus.\\
Recent measurements from $K^{+}_{e3}$~\cite{bnl865} and 
\ke~(KTeV~\cite{ktev-ckm}, KLOE~\cite{kloe} for $K_{S}$) are 
significantly above PDG values ''restoring'' unitarity, however their 
branching ratio are not in agreement with PDG averages. 
New experimental (and theoretical) inputs are therefore needed
to clarify the situation.

\section{ \kel~ Branching Ratio and Extraction of \vus }\label{sec:ke3br}

The determination of the $BR(K_{e3})$ proceeds through the measurement of the 
ratio $R$ of the decay rate of \ke~relative to all decays with two charged
particles in the final state ($2T$), namely $\pi e \nu$, $\pi e \mu$, 
$\pi^{+} \pi^{-} \pi^{0}$, $\pi^{+} \pi^{-}$, and 3 $\pi^{0}$ with Dalitz
decay of one $\pi^{0}$.
\be
R ~=~ \frac{\Gamma (K_{e3})}{\Gamma (2T)} ~=~ 
      \frac{N_{Ke3}/A_{Ke3}}{N_{2T}/A_{2T}}
\ee
$N_{i}$ and $A_{i}$ being the number of reconstructed events and the acceptance,
respectively. Since the $BR$s of the neutral channels have been measured, the sum 
of $BR$s of all \kl~decay modes with two charged tracks $BR(2T)$ 
is experimentally known: 
\be
BR(2T)= 1 - BR(All~Neutral) = 1.0048 - BR(3\pi^{0})
\ee
and $BR(K_{e3})$ can be derived from $R:~BR(K_{e3})=R \times BR(2T)$.
The \ke~events were selected using exactly the same selection as the $2T$ 
events with in  addition the requirement of $e$ identification from the LKr 
calorimeter.
This was done (see Fig.~1a) requiring $E/p > 0.93$ where $E$ is the energy 
released in the calorimeter and $p$ is the track momentum  measured in the 
spectrometer. 
The number of \ke~accepted events amounts to $6.8~10^{6}$.
The background from $K_{\mu 3}/K_{3\pi}$ events with a $\pi$ misidentified as
an electron was estimated from \ke~data with identified $e$ ($E/p >1$)
and is $5.8~10^{-3}$. The inefficiency of electron ID is determined from 
\ke~data with identified $\pi$ ($0.3<E/p<0.7$) and its value is 
$4.9~10^{-3}$. The detector response was reproduced using a GEANT based 
simulation. The average two track acceptance was obtained from a mean of 
the individual acceptances weighted with their BR. These were
evaluated combining together PDG and the recent KTeV~\cite{ktev-br}
results.
\begin{figure}
 \psfig{figure=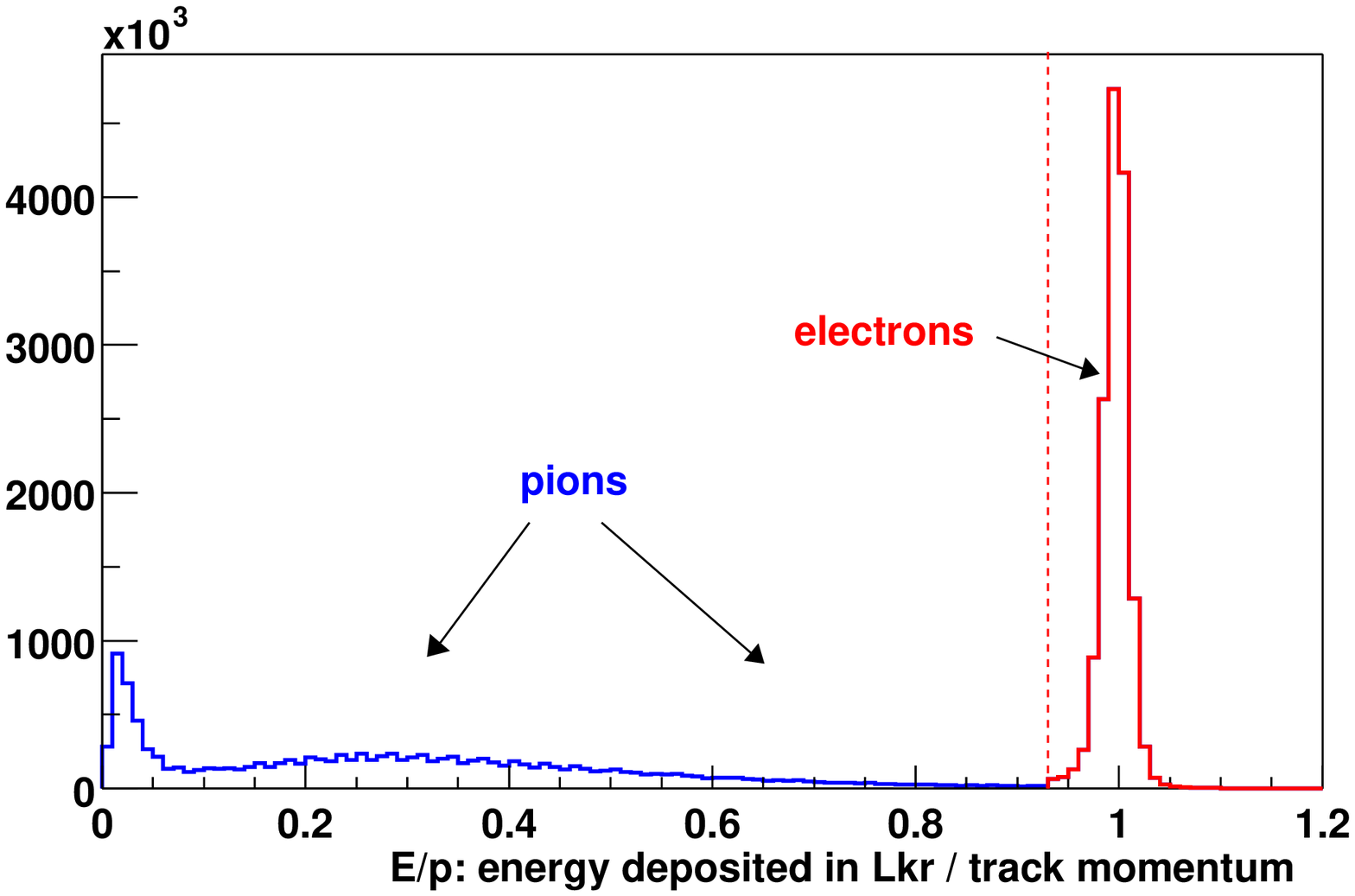,height=1.8in}
 \hskip 0.5 cm
 \psfig{figure=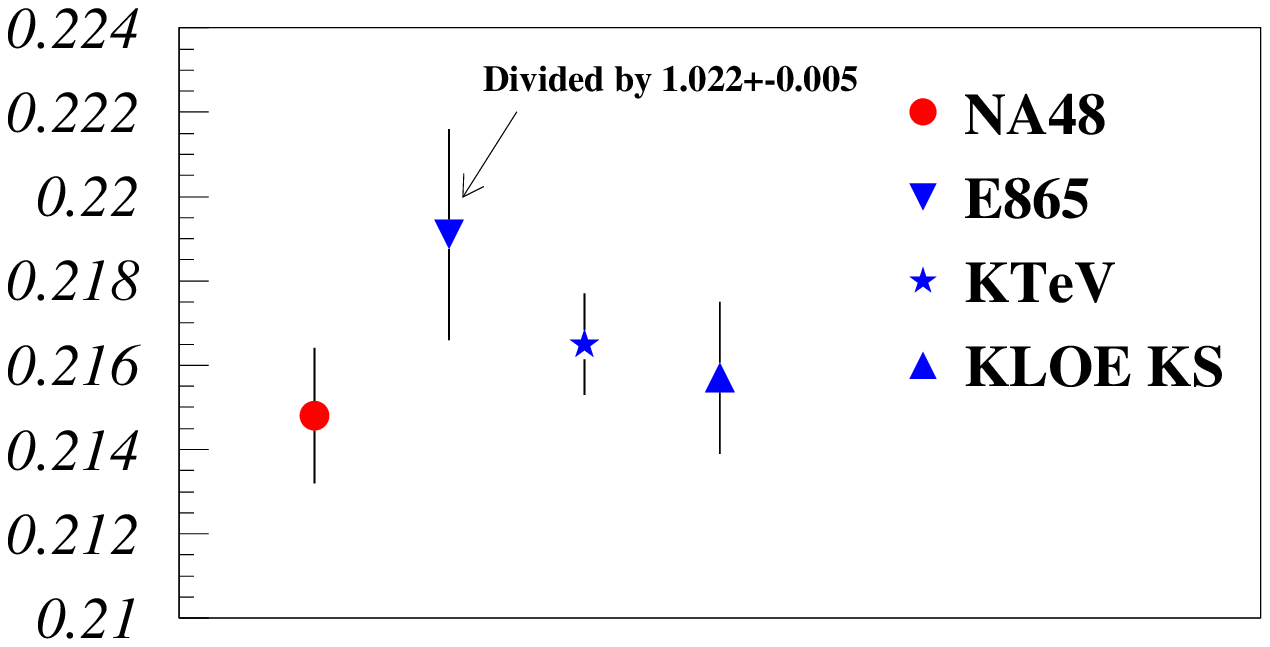,height=2.0in}
 \caption{(a) E/p distribution of selected \ke~events. 
          (b) Recent experimental results for \vus$f_{+}(0)$.  
 \label{figure1}}
\end{figure}
We obtained~\cite{na48-br} $R = 0.4975\pm 0.0035$ and 
$BR(K_{e3}) = 0.4010\pm 0.0028_{exp} \pm0.0035_{norm}$ where the first error is the 
complete experimental error and the second one is from the normalization.
To determine \vus~we followed the prescription of Ref.~\cite{cirigliano04}
obtaining \vus$f_{+}(0)=0.2146\pm0.0016$. Using their 
value of $f_{+}(0)=0.981\pm0.010 $ we get:
\be
\vert V_{\it us} \vert ~=~ 0.2187\pm0.0016_{exp} \pm 0.0023_{theo}
\ee
The experimental number of \vus$f_{+}(0)$ is in agreement with the recent
KTeV results~\cite{ktev-ckm} but larger than the PDG one. The result on 
\vus~is still 2.4~$\sigma$ lower than what required by CKM unitarity
and it is dominated by the theoretical uncertainties.

\section{\kel~Form Factors}
Assuming that only the vector coupling contributes to the decay, the matrix element
can be written in terms of two dimensionless form factors $f_{\pm}(t)$~:
\be
   {\frak M} = G_{F}/\sqrt{2}~ V_{us}~ \left[ f_{+}(t)
    \left(P_{K}+P_{\pi}\right)^{\mu}
    \bar u_{\ell} \gamma_{\mu} (1+\gamma_{5}) u_{\nu} +
    f_{-}(t) ~ m_{\ell} \bar u_{\ell} (1+\gamma_{5}) u_{\nu}  \right]
\ee
where $m_{\ell}$ is the lepton mass, $\bar u_{\ell}$ and $u_{\nu}$ are the
lepton spinors and $P_{K}$ and $P_{\pi}$ are the kaon and pion four--momenta 
respectively. 
Since the contribution of $f_{-}$  is proportional to the lepton mass squared
it can be neglected in \ke~decays.
The determination of the form factor is based on the measurement 
of the Dalitz plot density:
\be
  \rho(E_{e}^{*},E_{\pi}^{*}) =
\frac  { dN^2 (E_{e}^{*},E_{\pi}^{*})} { dE_{e}^{*}~ dE_{\pi}^{*} }
  \propto f_{+}^{2}(t)~A
\ee
$A$ is a kinematical term and $E_{e}^{*}$ and $E_{\pi}^{*}$ are the electron
and pion energies in the kaon c.m., respectively. An usual assumption is that
the form factor depends lineraly on $t$ (the square of the four--momentum 
transferred to the lepton system):  
$ f_{+}(t) =  f_{+}(0) ( 1 + \lambda_{+} t/m_{\pi}^{2} )$. \\
A fit comparing the data and the MC expectations was performed in order 
to extract the form factors. The results, allowing also for 
scalar ($f_{S}$) and tensor ($f_{T}$) contributions to the decay, are:
\vskip -0.5 cm

\begin{equation}
\begin{array}{rcl}
 \rule{0mm}{5mm} \lambda_{+} & = &0.0284\pm0.0007_{stat}\pm 0.0013_{syst} \\
 \rule{0mm}{5mm} \vert \frac{ f_{S}}{f_{+}(0)} \vert & = & 
                             0.015^{+0.007}_{-0.010~stat}\pm0.012_{syst} \\
 \rule{0mm}{5mm} \vert \frac{ f_{T}}{f_{+}(0)} \vert & = & 
                             0.05^{+0.03}_{-0.04~stat}\pm 0.03_{syst} \\
\end{array}
\end{equation}
The results do not support any evidence for scalar or tensor couplings,
if the fit is done assuming only the vector contribution one gets:
\be
 \lambda_{+} ~=~  0.0288\pm 0.0005_{stat}\pm 0.0011_{syst}
\ee
Figure 2b shows the comparison between the experimental 
results for the slope $ \lambda_{+}$.

\begin{figure}
\psfig{figure=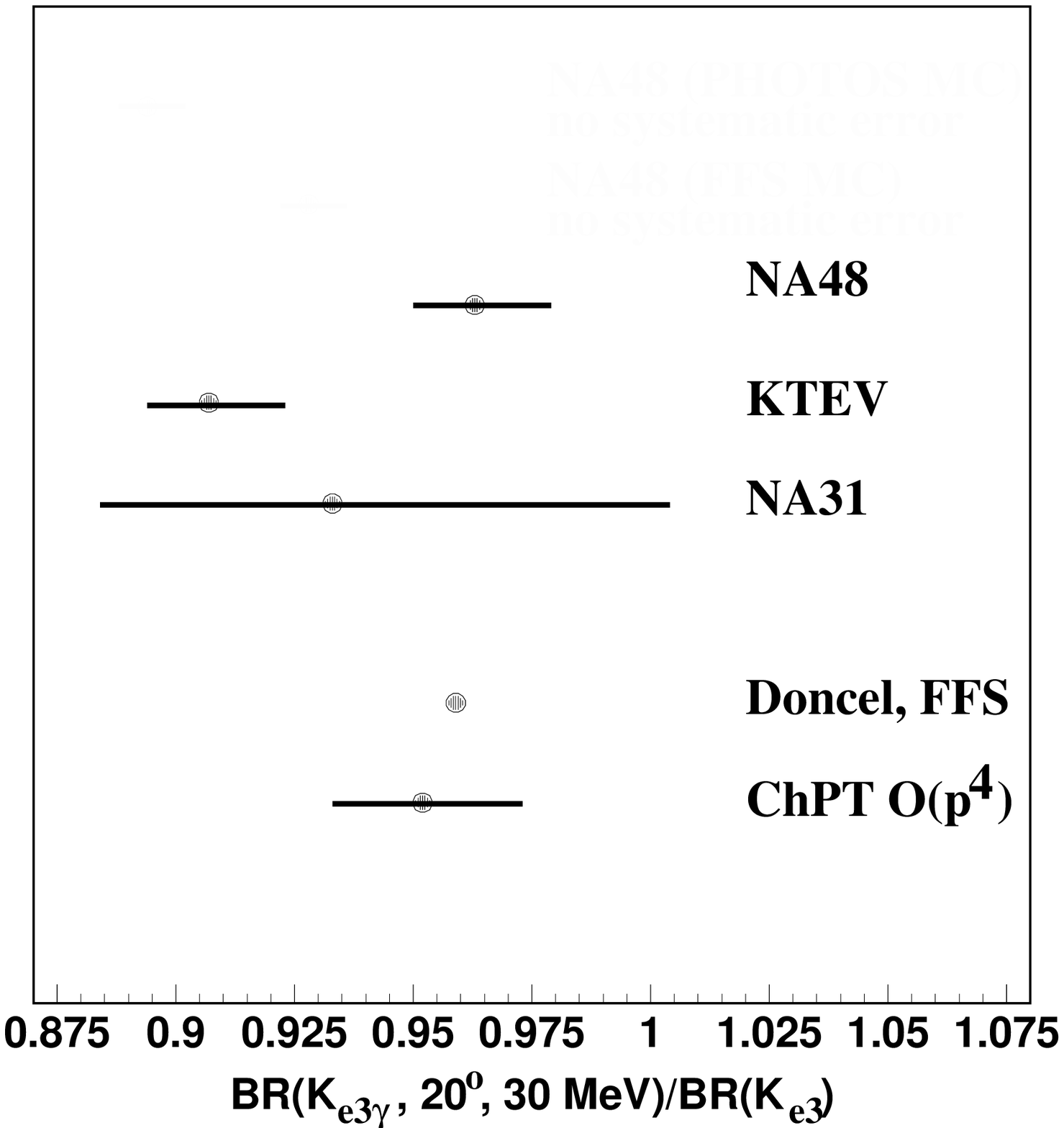,height=1.8in,width=2.5in}
\hskip 1. cm
\psfig{figure=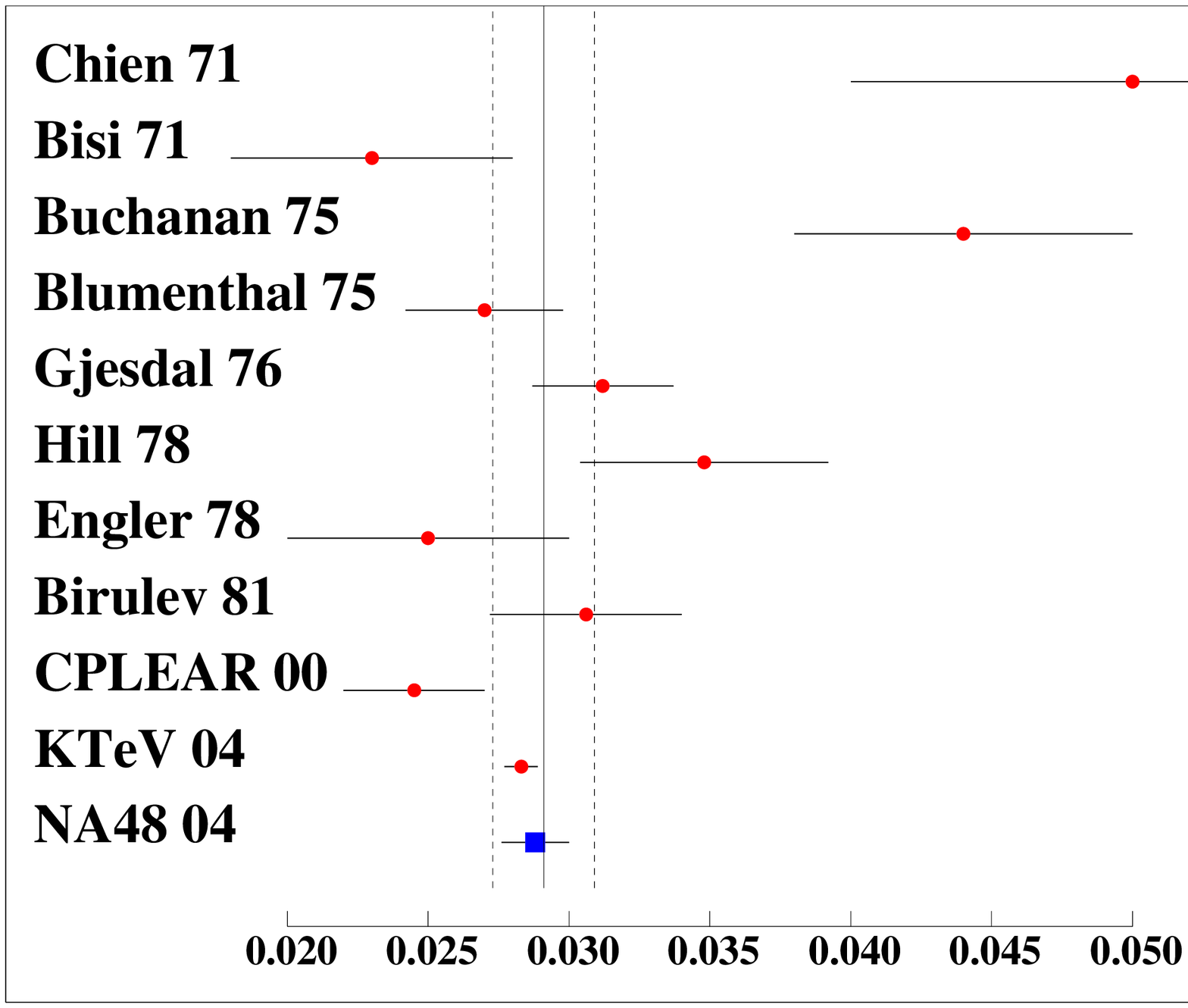,height=1.8in,width=2.8in}
\caption{(a)~Comparison between the measured values
and theoretical predictions of the ratio $BR$(\keg)/$BR$(\ke). 
(b)~Experimental results for the slope  $\lambda_{+}$ of \ke~decays. 
\label{figure2}}
\end{figure}

\section{\kegl~Branching Ratio}
In total 18977 \kegl~events have been reconstructed from the data sample.
The events are selected in the same manner as the \ke~events adding the 
requirement of exactly one $\gamma$ in the LKr calorimeter. 
To take into account the effect of the radiative corrections (real and  
virtual) the Monte Carlo events, generated with the  PHOTOS~\cite{photos} 
package, were modified in order to reproduce the data. This was achieved by
weighting the MC events with the distribution, obtained from the data, of 
$\theta^{*}_{e\gamma}$ the angle between the electron and the photon in 
the kaon c.m. In this way the agreement between data and MC has been found 
very good for all the variables. The preliminary result of the branching 
ratio measurement is:
\be
\frac {BR(K_{L}\rightarrow~\pi^{\pm}~e^{\mp}~\nu_{e} \gamma, 
~E^{*}_{\gamma} > 30 MeV,~\theta^{*}_{e\gamma} > 20^{\circ}) }
{ BR(K_{e3}) } ~=~
(0.964\pm0.008^{+0.012}_{-0.011})\%
\ee
As it can be seen in Fig.~2a, this is in good agreement with  theoretical 
predictions~\cite{FFS,doncel} and recent ChPT 
calculations~\cite{ke3g-chpt1,ke3g-chpt2} and in disagreement with the other high 
precision measurement done by the KTeV collaboration~\cite{ktev-ke3g}.

\end{document}